\documentclass[aps,pra,reprint,amsfonts,amssymb,amsmath,nofootinbib]{revtex4-2}

\usepackage{siunitx}% Unify displaying of units
\usepackage{graphicx}% Include figure files
\usepackage{dcolumn}% Align table columns on decimal point
\usepackage{hyperref}% add hypertext capabilities
\hypersetup{hidelinks}
\usepackage{cleveref}% Unify sec/fig/table refs
\usepackage[normalem]{ulem}
\usepackage{tikz}
\usepackage{bm}
\Crefname{equation}{Eq.}{Eqs.}
\Crefname{figure}{Fig.}{Figs.}
\Crefname{section}{Sec.}{Sections}

\begin{document}
%\title{Harmonizing scalar field dark matter searches with gravitational-wave detectors using musical tools}
% \title{Advancing real scalar field dark matter searches\\in Fabry-Pérot Michelson interferometers}
\title{Fast and Precise Spectral Analysis for Dark Matter Searches with LIGO}
%\thanks{A footnote to the article title}

\author{Alexandre S. Göttel$^{1}$\thanks{Lorem Ipsum}}
\email[Correspondence email address: ]{gottela@cardiff.ac.uk}
\author{Vivien Raymond$^1$}
\affiliation{$^1$Gravity Exploration Institute, Cardiff University, Cardiff CF24 3AA, United Kingdom}
\date{\today} % Leave empty to omit a date

\begin{abstract}
We introduce a novel logarithmic spectral estimation method for dark matter searches using gravitational-wave detectors, integrating established dark matter search techniques with insights from computer music analysis. By leveraging symmetries between the time and frequency domains, this method matches the computational efficiency of FFT-based algorithms without, unlike such algorithms, compromising precision. We apply this approach to data from LIGO’s third observing run, directly comparing its performance with that of a previous search. Our results show a consistent $15\%$ improvement across nearly the entire frequency range, without additional computational costs. With potential for further refinements, this method already offers a solution capable of maximizing the scientific potential of current and future gravitational-wave observatories.
\end{abstract}

\maketitle
% - ConstQ approximation
%   - Elucidate SNR maximisation
%   - Talk about block approx first in reference to Search paper
%   - Explain idea behind approx
%   - Go into Kaiser-Bessel-specifics
% - Mention HDF+Mem improvements? Not sure..
% - Data selection
%   - Optimisation for additional freq. blocks
%   - Don't force overlaps!
% - Analysis details:
%   - Building iterative background models
%   - Likelihood including calibration marginalisation
%   - Candidate treatment and peak shape derivation
%   - Include statistical fluctuation of DM effect?
% - Experimental changes:
%   - Details of TF calculation
%   - Light travel time
%   - De Broglie effect (segment-specific sky averaging?)
Theoretical candidates for Dark Matter (DM) span a broad spectrum, from ultralight ($10^{-22}~\SI{}{\electronvolt}$) particles to sub-solar-mass primordial black holes. Some of these candidates, assuming a non-zero coupling with standard model fields, would produce detectable signals in Fabry-Pérot Michelson interferometers like the gravitational-wave observatories LIGO, Virgo and KAGRA~\cite{collaboration2015,acernese2014,KAGRA}. These interactions range from altering physical constants~\cite{grote2019a}, to modifying the behaviour of laser beams~\cite{nagano2019,Nagano:2021kwx}, accelerating optical components~\cite{pierce2018a,Guo:2019ker,KagraAuxiliary}, and affecting existing gravitational-wave signals~\cite{Nguyen:2024fpq}, or even being detectable through gravitational interactions~\cite{Manita:2022tkl,Lee:2022tsw,Armaleo:2020efr}.
Recently, these aspects have been used in studies that exploit the extreme sensitivity of laser interferometers to set upper limits on various DM candidates~\cite{abbott2022,LIGOVectorDM,vermeulen2021a}. This advances our understanding of DM while highlighting the versatility of gravitational-wave detectors in exploring physics beyond the standard model.

\noindent Particularly, we have shown in~\cite{goettel2024} that LIGO can outperform any other direct scalar DM search in a mass band from roughly $10^{-14}~\SI{}{\electronvolt}$ to $10^{-11}~\SI{}{\electronvolt}$, with results only about two orders of magnitude away from indirect ``fifth-force'' constraints~\cite{hees2018,berge_microscopes_2021}. However, all previous interferometer-based searches have had to either tackle significant computational challenges or sacrifice some level of sensitivity. This study introduces a new method, which for the first time is able to bring down computational cost by orders of magnitude without resorting to any approximations. We demonstrate that this new approach is key to fully leveraging future detectors, and, importantly, we highlight that it is broadly applicable to all similar DM searches (see~\cite{morisaki2021,abbott2022,LIGOVectorDM,Miller:2020vsl,Miller:2023kkd,Miller:2022wxu}).
\Cref{sec:theory} introduces the target DM interactions and the detector couplings considered, \Cref{sec:maths} details the challenges facing this data analysis and methods to solve them, \Cref{sec:strategy} establishes a unified statistical framework used to derive our upper limits, and \Cref{sec:results} evaluates and compares our results to previously-used methods.

% \subsection*{Scalar field dark matter in LIGO}
\section{Scalar field dark matter in LIGO}  \label{sec:theory}
% Theory (small!)
As discussed in~\cite{goettel2024}, scalar field Dark Matter originating in the early universe would behave today as a coherently oscillating classical field~\cite{arvanitaki2015,stadnik2015}:
\begin{equation}\label{eq:osc_field}
        \phi(t,\vec{r}) = \phi_0 \cos\left(\omega_\phi t - \vec{k}_\phi \cdot \vec{r}\right),
\end{equation}
where $\omega_\phi$ is the Compton frequency of the DM, and $\vec{k}_\phi$ its wave vector. Assuming that the field is responsible for the local DM density $\rho_{\rm local}$, the amplitude is given by $\phi_0 = \sqrt{2\rho_{\rm local}}/m_\phi$.

\noindent Following~\cite{ringwald2012,hees2018}, we consider linear couplings to the standard model Lagrangian:
\begin{equation}\label{eq:L_int}
    \mathcal{L}_\mathrm{int} \supset \frac{\phi}{\Lambda_\gamma} \frac{F_{\mu\nu}F^{\mu\nu}}{4}  - \frac{\phi}{\Lambda_e} m_e \bar{\psi}_e \psi_e,
\end{equation}
where $\psi_e$, $\bar{\psi}_e$ are the SM electron field and its Dirac conjugate, respectively, $F_{\mu\nu}$ is the electromagnetic field tensor, $m_e$ is the electron rest mass, and $\Lambda_\gamma$, $\Lambda_e$ represent the couplings.

\noindent Since LIGO's calibrated data is provided as \textit{gravitational-wave-induced} strain, but our analysis requires a strain that reflects the interferometer's response to \textit{DM-induced} oscillations, we follow the procedure outlined in~\cite{goettel2024} to map LIGO data to DM coupling constants. This yields:
\begin{equation} \label{eq:DM_Lambda}
h(\omega)\cdot A_{\rm cal}(\omega) \approx \left(\frac{1}{\Lambda_\gamma} + \frac{1}{\Lambda_e}\right) \cdot\frac{\hbar\,\sqrt{2\,\rho_{\rm local}}}{m_\phi\, c}, \\
\end{equation}
where $h(\omega)$ is the \textit{GW-induced} strain at a frequency $\omega$, and $m_\phi$ is the Compton mass of DM. $A_{\rm cal}$ is a conversion factor containing transfer function information which was obtained through detailed optical simulations~\cite{goettel2024}. We note that these simulations automatically include effects stemming from a finite light travel time, as discussed in~\cite{TravelTime,Manita:2023mnc}

% Use Gaia RD2 229 km/s -> (2.917 +- 0.013) 10^-7 relative bin width
% Coherence time = 10^6 / \omega_\phi (derevianko)
% FWHM delta_omega = 2.5/tau_c ~ 7.3e-7 f
\noindent Assuming that dark matter is virialised as in a standard galactic halo model, its velocity distribution is expected to follow a Maxwell-Boltzman distribution:
\begin{equation}
    f(\vec{v}) = \sqrt{\frac{2}{\pi}}\frac{|v|^2}{\sigma_v^3}e^{-\frac{1}{2}\frac{|v|^2}{\sigma_v^2}},
\end{equation}
where $\sigma_v = \Theta_0\cdot\sqrt{3/2}$ is the characteristic virial dispersion velocity and $\Theta_0 = \SI[separate-uncertainty]{229\pm 1}{\kilo\meter\per\second}$~\cite{catena2012,cautun2020}. Practically, we model the DM as isotropic and homogeneous in the galactic halo, with a cutoff at the escape velocity $v_{\text{esc}} = \SI[separate-uncertainty]{528\pm 25}{\kilo\meter\per\second}$~\cite{deason2019}.
The observed DM frequency $\omega_{\text{obs}} = \omega_\phi + \frac{1}{2}m_\phi\vec{v}_{\text{obs}}^2$ is expected to have a Doppler-broadened linewidth.
With the solar system moving at the galactic circular velocity $\Theta_0$ and taking Earth's orbit into account, we estimate $\delta_f/f \approx 10^{-6}$, where $\delta_f$ is the linewidth of the Doppler-shifted peak.

% Further, assuming that the DM is trapped and virialized in our galaxy through a standard halo model, the observed frequency $\omega_{\text{obs}} = \omega_\phi + \frac{1}{2}m_\phi\vec{v}_{\text{obs}}^2$ is expected to have a Doppler-broadened linewidth.

\noindent This results in a finite coherence time $\tau_c = (\sigma_v/c)^2/\omega_\phi$ for the DM signal~\cite{derevianko2018}. Any frequency-space search will thus have to integrate precisely over this coherence time to avoid losing signal contributions while minimising noise~\cite{vermeulen2021a}. We achieve this using a Fourier transform, but since $\tau_c$ is frequency-dependent this cannot be performed with the Fast Fourier Transform (FFT)~\cite{cooley1965} algorithm.
This represents a significant technical challenge, especially given the coherence times reached at our lowest explored frequencies ($\SI{10}{\hertz}$, corresponding to $\tau_c \approx \SI{28}{\hour}$), because without access to the speed of the FFT, costs scale with $\mathcal{O}(N^2)$, where $N$ is the number of strain data points, and quickly enter a prohibitively expensive regime.

\section{Methods} \label{sec:maths}
\noindent We hereinafter use Logarithmic Power Spectral Density (LPSD) nomenclature from~\cite{troebs2006}. As such, we can re-write the equation for a discrete Fourier transform (DFT) as:
\begin{equation} \label{eq:DFT}
    X_j = \sum_{n=0}^{N(j)-1} x_n w_{j,n} e^{-2\pi in\mathcal{Q}/N(j)},
\end{equation}
where $X_j$ is the Fourier coefficient in the $j$-th frequency bin, the index $n$ sums over time, $x$ represents our data, $w_j$ is a window function, and $N(j)$ is the frequency-dependent number of data points associated with the coherence time $\tau_c$. We used:
\begin{equation}
    \mathcal{Q} = f / \delta_f \approx 10^6,
\end{equation}
which, being constant, encapsulates the frequency dependence of the exponential term through $N(j)$, highlighting the impossibility of applying a standard FFT.
Assuming that a frequency axis is made of $J$ logarithmically-spaced bins between the analysis' minimum and maximum frequencies f$_{min}$ and f$_{max}$:
\begin{align} \label{eq:fj}
    %&\log \text{f}_k = \log \text{f}_{min} + \frac{k}{J - 1}(\log \text{f}_{max} - \log \text{f}_{min})\\
    &\text{f}(j) = \text{f}_{min}e^{\frac{j}{J - 1}(\log \text{f}_{max} - \log \text{f}_{min})},
\end{align}
where $f(j)$ is the analysed frequency in bin $j$. Since, by definition, the frequency spacing $\text{f}_{j+1} - \text{f}_j = \text{f}_j / \mathcal{Q}$, we further arrive at:
\begin{equation} \label{eq:J}
    J = 1 + \frac{\log \text{f}_{max} - \log \text{f}_{min}}{\log (1 + \mathcal{Q}) - \log (\mathcal{Q})}.
\end{equation}
Unsurprisingly, $J$ depends only on the width of the frequency axis and our resolution term $\mathcal{Q}$. Finally, the discrete integration length $N(j)$ (corresponding to our coherence time) can simply be defined as $\text{f}_s / (\text{f}_{j+1} - \text{f}_j)$, which, using \Cref{eq:fj,eq:J}, leads to:
\begin{align} \label{eq:segment_length}
    %N(k) &= \text{f}_s / (\text{f}_{k+1} - \text{f}_k)\\
    N(j) &= \frac{\text{f}_s}{\text{f}_{min}}\mathcal{Q}\cdot\left(\frac{\mathcal{Q}}{1 + \mathcal{Q}}\right)^j,
\end{align}
where f$_s$ is the sampling frequency.

\noindent We follow the approach used in~\cite{troebs2006,goettel2024} for the power spectrum, whose calculation summarises to:
\begin{equation} \label{eq:PSD}
    A_j^{total} = \frac{2}{f_s}\frac{\frac{1}{N_s(j)}\sum_{s=0}^{N_s(j)-1} |X_{j, s}|^2}{\sum_{n=0}^{N(j)-1} w_n^2},
\end{equation}
where $A_j^{total}$ is the power spectral density in the frequency bin $j$, the power $X_j$ is averaged over data sections, labelled with the subscript $s$, following the Welch method~\cite{welch1967}, $N_s(j)$ is the number of data sections, and $w_n$ is a window function. As one can see, the term in the numerator is simply an average of the power, while the denominator is a frequency-dependent normalisation term. To help visualise the scale of the analysis, a list of commonly used variables and their values in this study can be found in \Cref{tab:variables}.

%While a full analysis on real data will be described below, it is worth mentioning here that for practical reasons we individually analyse separate data segments of at least $N(0)$. Within the software, each data segment is then further divided in frequency-dependent sections of length $N(j)$, which is required for the DFT calculations (see \Cref{eq:DFT}) to be valid. Evidently, at high frequencies where $N(j)$ is much smaller than $N(0)$, several such sections can fit inside of the full segment. In order to make use of all of the available segmented data, we thus repeat our calculations over each section and average our results over them.

\subsection{Band approximation} \label{sec:bandApprox}
As seen above, the dependence of $N(j)$ on frequency prohibits the use of a simple FFT. However, it is also clear from \Cref{eq:segment_length} that the frequency-dependence is quite slow ($\propto (1 + \mathcal{Q}^{-1})^j$, where $\mathcal{Q} \gg 1$). A natural solution to those computational problems is thus to make approximations related to $N$. This idea was used before, \textit{e.g.}~\cite{Miller:2020vsl,piccinni2018} where the data was split in \SI{10}{\hertz}-wide frequency bands for FFTs, and in~\cite{goettel2024} we instead defined the length of each frequency band by requiring that the deviation from the optimal coherence time remains within a certain fraction $\epsilon$. Specifically, using \Cref{eq:segment_length} we set:
\begin{align} \label{eq:deltaj}
    &N(j+\delta j) \leq (1 - \epsilon) N(j)\\
    \leftrightarrow\ &\delta_j^\text{max} = \frac{\log(1 - \epsilon)}{\log(\mathcal{Q}) - \log(1 + \mathcal{Q})}.
\end{align}
With LIGO, $\delta_j^\text{max} \approx 10^4$ when $\epsilon = 1\%$. This very slow variation of $N(j)$ over frequency thus allows us to simply set $N$ to be constant over frequency bands of (up to) $\delta_j^\text{max}$ bins. Defining these blocks in terms $\delta_j^\text{max}$, as opposed to a specific width in \SI{}{\hertz}, ensures that the deviation from the optimum never exceed the threshold set by $\epsilon$, which is equivalent to a threshold in SNR loss.

\begin{figure}[h]
    \centering
    \includegraphics[width=.5\textwidth]{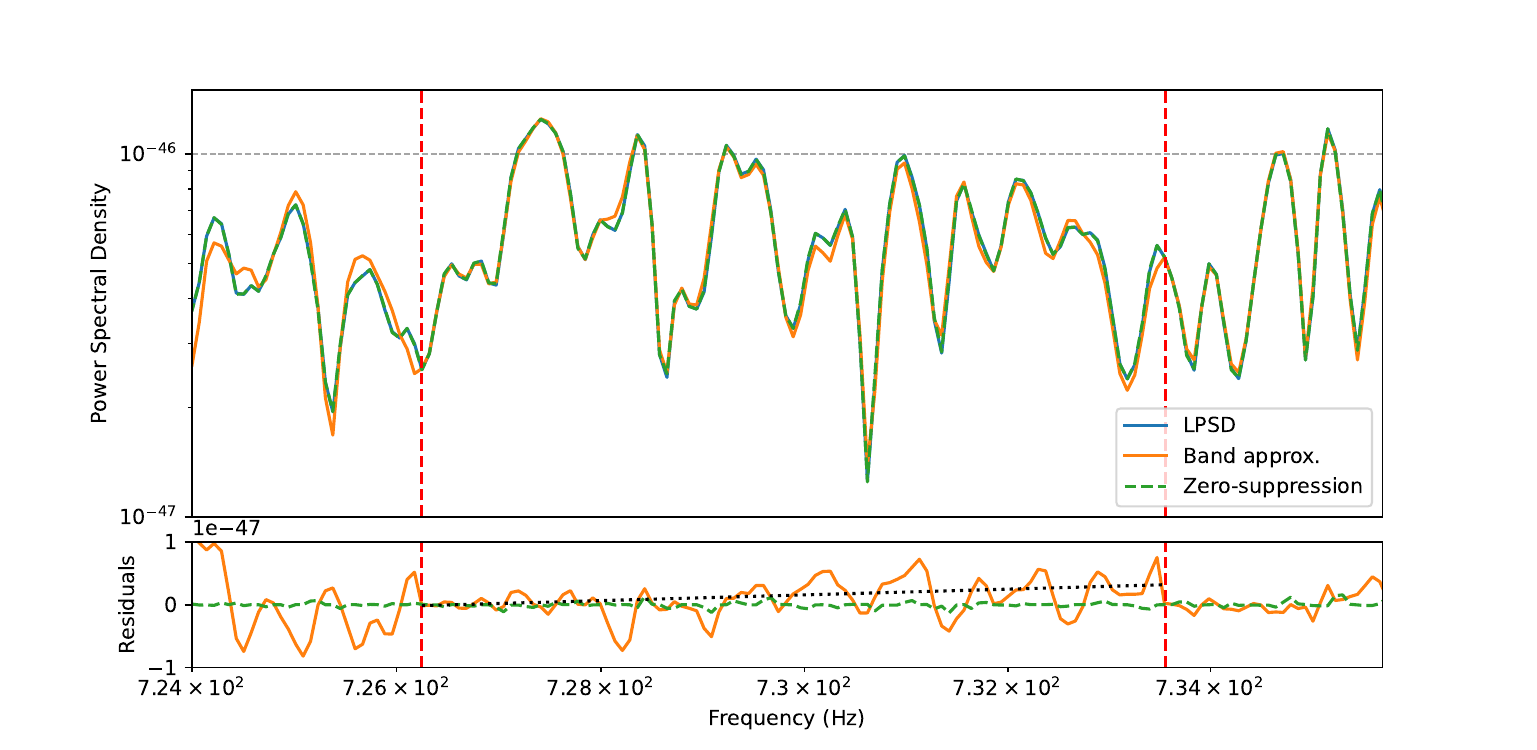}
    \caption{PSD and residuals between methods: the solid orange lines depict band approximation results ($\epsilon = 1\%$), the dashed lines the zero-suppression results. The reference solid blue line (LPSD) from a brute-force calculation is fully obscured by the dashed line, showing, as reflected in the residuals, how precise the zero-suppression is. A full band, roughly between \SIrange{726}{733}{\hertz}, is marked by vertical lines. A gradual \textit{drift} is exemplified by the dotted line, resulting from a least-squares fit to the residuals from the band approximation results.}
    \label{fig:block_results}
\end{figure}

\noindent As demonstrated in~\cite{goettel2024}, this method can be used competitively while providing a $\mathcal{O}(10^4)$ speed-up. However, it does have a few drawbacks, one of them naturally being the sub-optimal integration time - though the overall SNR loss is small~\cite{vermeulen2021a}. However, as seen in \Cref{fig:block_results}, the structure also induces a \textit{drift} in frequency as the sub-optimal $N(j)$ diverts from its corresponding $\tau_c$, which must be corrected for in an additional post-processing step.

\subsection{Spectral zero-suppression} \label{sec:constQ}
\begin{figure}
    \centering
    \includegraphics[width=.5\textwidth]{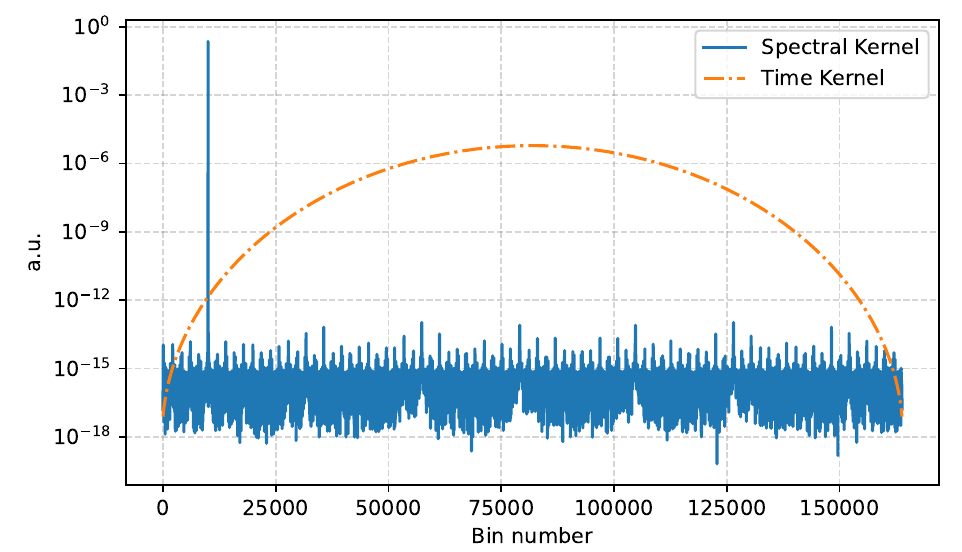}
    \caption{Absolute value of a time kernel (dot-dashed) and its associated spectral kernel (solid) for an example frequency. As one can see, the spectral kernel is much more sharply peaked, to the point where most values (scattered around $10^{-15}$) are dominated by float precision uncertainty.}
    \label{fig:time_v_kernel}
\end{figure}

This section introduces an approach based on that developed (originally for computer-music applications) in~\cite{brown1992a}. The basic idea is to identify the terms in \Cref{eq:DFT} with a data term $x$ and a \textit{kernel} $\mathcal{K}$:
\begin{equation} \label{eq:defineKernel}
    X_j = \sum_{n=0}^{N(j)-1} x_n w_n e^{-2\pi in\mathcal{Q}/N(j)} = \sum_{n=0}^{N(j)-1} x_n\mathcal{K}^*_n,
\end{equation}
where the asterisk denotes the complex conjugate, in order to make use of Parseval's theorem:
\begin{equation} \label{eq:parseval}
    \sum_{n=0}^{N-1} x_n\mathcal{K}^*_n = \frac{1}{N}\sum_{k=0}^{N-1}\Tilde{X}_k\Tilde{K}_k^*,
\end{equation}
where $\Tilde{X}$ and $\Tilde{K}$ represent the Fourier transforms of $x$ and $\mathcal{K}$, respectively. The reason behind using this identity becomes apparent in \Cref{fig:time_v_kernel}, where one can see that while the kernel is very broad in the time domain, it is very sharply peaked in the frequency domain. Indeed, in the context of a LIGO dark matter search, the lowest analysed frequency (\SI{10}{\hertz}) is associated with a DFT length of $N(0) \approx 1.7\cdot 10^9$, but the number of non-zero bins in the spectral kernel is about ten (where we define zero-bins as those contributing negligibly to the sum in \Cref{eq:parseval}). Therefore, a simple zero-suppression on the spectral calculation can achieve a speed-up of $\mathcal{O}(10^8)$ compared to the time-domain operation. The ratio between the DFT length and the number of significant bins in the spectral kernel, and therefore the achievable speed-up, decreases exponentially as the frequency (or, equivalently, $k$) increases. Fortunately, even at the highest frequencies considered here (\SI{5000}{\hertz}) however, the ratio is still around 100. Taking only this relationship into account, one can estimate the speed-up through zero-suppression for an analysis across our entire frequency range from \SIrange{10}{5000}{\hertz} to be about $2\cdot10^4$.

\noindent This number does not take into account the cost of calculating $\Tilde{X}_k$ and $\Tilde{K}_k^*$. In~\cite{brown1992a}, the authors suggest to calculate the latter in advance, by performing an FFT on the time-domain kernel for every frequency bin, so that it can be simply applied to the data later. In musical applications, where one rarely considers more than a few thousand frequency bins, this will indeed result in an effective speed-up because the kernel only needs to be calculated once. However in our case, with a number of frequency bins well exceeding $10^6$, one can easily see that the cost of calculating the kernel in this way will be at least an order of magnitude more costly than the original DFT calculation we wish to speed-up (\Cref{eq:DFT}).
%Since the number data of segments we analyse is less than 100, the effective speed-up would at best be quite low.

\noindent Similarly for $\Tilde{X}_k^*$, performing a Fourier transform of the data for every frequency bin, even if using an FFT, would negate any possible speed-up. We instead derive a fully analytical solution for the spectral kernel, rendering the calculation of $\Tilde{K}_k^*$ effectively ``free'' computationally. Moreover, since \Cref{eq:parseval} applies to any Fourier transform with consistent lengths for both $x$ and $\mathcal{K}$, we can allow a single $\Tilde{X}_k$ to be reused across all frequency bins by simulating zero-padding in our analytical solution for $\Tilde{K}_k^*$. More detailed explanations as well as a full derivation can be found in \Cref{app:analyticalKernel}.

\noindent In summary, using this approach along with our analytical framework allows us to calculate all $X_j$ terms (see \Cref{eq:PSD}) with a single FFT of the data followed by a heavily zero-suppressed transformation through spectral domain operations. While in theory this constitutes an approximation, because none of the suppressed terms are exactly zero, in practice the difference is negligible as we set a stringent relative threshold of $10^{-15}$ on whether or not to consider values to be zero. This approach is thus precise, does not require any additional post-processing (unlike the approach described in \Cref{sec:bandApprox}), and almost fully recovers the $\mathcal{O}(N\log N)$ scaling power of the FFT.

\noindent Following those improvements, the computational bottleneck in the PSD calculation shifts to the evaluation of the frequency-dependent normalisation term (see the denominator in \Cref{eq:PSD}). This is because the (Kaiser) window function we use involves computationally expensive calls to Bessel functions. Closer inspection (see \Cref{app:normalisation}) revealed that the whole term is proportional to the segment length ($N(j)$, see \Cref{eq:segment_length}). This allowed us to further reduce the computational cost drastically by computing the normalisation term once and scaling it via simple multiplication for all other frequency bins.

\begin{table}[h]
    \centering
    \begin{tabular}{c|c}
        Variable & Value \\ \hline
        $J$ & 6214612 \\
        $f_{min}$ & \SI{10}{\hertz} \\
        $f_{max}$ & \SI{5000}{\hertz} \\
        $f_s$ & \SI{16384}{\hertz} \\
        $\epsilon$ & 0.01 \\
        $\mathcal{Q}$ & $10^6$
    \end{tabular}
    \caption{Commonly referred variables and their values in this analysis.}
    \label{tab:variables}
\end{table}

\subsection{Search strategy} \label{sec:strategy}
\begin{figure}
    \centering
    \includegraphics[width=\linewidth]{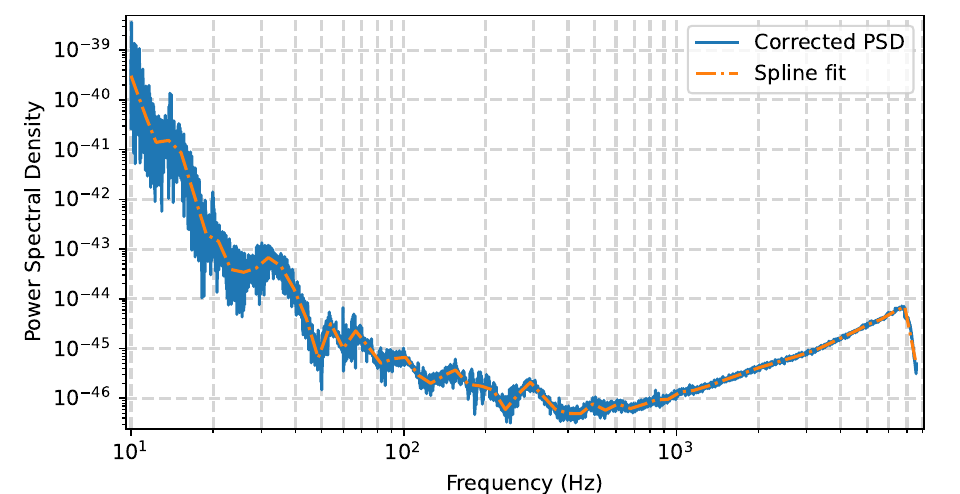}
    \caption{Example LIGO power spectral density (PSD), after removal of excess lines found by the procedure described in the text. The dot-dashed line shows the results of the spline fit from said procedure. These fit results, with the variance and skewness of the remaining PSD data (solid line), define our background model.}
    \label{fig:example_PSD}
\end{figure}

Having calculated PSDs using \Cref{eq:PSD} and both aforementioned methods, our search for dark matter becomes a search for localised excess in power spectral density. However, since the effect from DM cannot be ``turned off'', it is necessary to start by building a background model that is resistant to existing peaks in the data. Similar to methods used by LIGO calibration~\cite{sun2020}, we find that a cubic spline in log-log space can well describe the PSD as a function of frequency, see for reference \Cref{fig:example_PSD}.
We further find empirically that the residuals between the log of the PSD and said splines can be well described locally by skew normal (SN) distributions.

\begin{figure}[h]
    \centering
    \includegraphics[width=\linewidth]{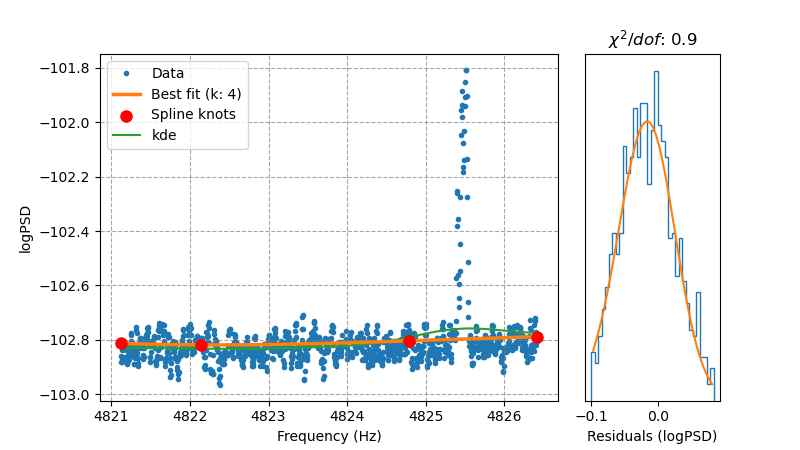}
    \caption{Left) The blue data points represent log(PSD) values as a function of frequency. The green dashed line shows a kernel density estimation (KDE) over these values. The solid orange line indicates the spline fit, with individual knots marked by large red dots. right) distribution of the residuals with accompanying fitted skew normal distribution. As one can see in this example, the combination of spline fits and skew normal distributions can well describe the data while remaining undisturbed by the presence of outliers in the PSD. The difference between KDE and Spline at the lower end of the frequency axis stems from the former not taking the skewness of the distribution into account, and the overestimation on the right side is ``leakage'' from the peak.}
    \label{fig:PSD_fit}
\end{figure}

\noindent In order to describe the background, one could in theory perform a maximum likelihood fit through an entire PSD. This however encounters a few problems. First is the high number of required variables resulting in an unstable procedure, second is the effect of (known and unknown) so-called lines in the PSDs which would affect naive fitting attempts, and third is the large variation in the properties of LIGO PSDs as a function of frequency. We solve all three issues by implementing a localized procedure in which the PSD is divided into windows of only 1000 frequency bins. Each fit is performed individually within such a window, though window bounds are set to overlap to ensure consistency around boundaries. In each such window, where the SN parameters can safely be viewed as constant, we fit a spline by minimising a skew normal likelihood through the residuals, see \Cref{fig:PSD_fit}. By combining the results from all windows into a single background model, we are able to capture the complexity and frequency-variation of LIGO PSDs. To minimise the effect of existing peaks (presumably including dark matter effects) on the results of those fits, we employ an iterative procedure for each window:
\begin{enumerate}
    \small
    \item Initialise the spline fit with $k = 4$ knots.
    \item Minimise the skew normal likelihood.
    \item Build a histogram of the residuals and discard all data in bins with heights below 5\% of the maximum bin height.
    \item Minimise the likelihood again.
    \item Compute the Bayesian Information Criterion (BIC).
    \item Increase $k$ by 1 and repeat steps 2 to 5 until the BIC stops increasing.
\end{enumerate}

\noindent The removal of low-contributing bins ensures robustness of the fitted background SN parameters against peaks. The Bayesian part of this approach also prevents overfitting, allowing us, as shown in \Cref{fig:PSD_fit}, to reconstruct the overall background shape of the data.
% Fitting the SN distribution with binned values also lets us calculate a $\chi^2$ value.

\noindent When using the band approximation (see \Cref{sec:bandApprox}), an extra step is required. Since the band locations (see \Cref{eq:deltaj}) are known a-priori and because the PSD \textit{drift} (see \Cref{fig:block_results}) scales linearly with frequency, we can simply add lines to the spline fit model. To treat those lines consistently over the frequency axis, one must take care to fit simultaneously over at least an entire band - typically including several windows. Finally, we set a cut on the quality of the fits by using the aforementioned histograms to calculate a $\chi^2$ value (assuming Poisson uncertainties). We note that this value is not, like the results of a proper $\chi^2$ fit, statistically expected to fluctuate around the number of degrees of freedom, but we find it is still a good proxy to indicate \textit{goodness-of-fit}. We exclude all data from windows with a corresponding $\chi^2$ value below 0.5 or above 10.
%Only a few windows fail this criterion, and, as shown in \Cref{sec:results}, since this analysis is performed on 40 data segments, there are no gaps in the likelihood as defined below.

\noindent To combine data from different time segments and interferometers, which would otherwise be challenging due to variations in detector response to DM effects and calibration, we adopt the likelihood-based approach outlined in~\cite{goettel2024}, where detector responses were modelled using simulated transfer functions. The resulting likelihood is defined as:
\begin{equation} \label{eq:lkl}
    \begin{aligned}
        \log&\mathcal{L}(\mu, \bm{\theta}, \tilde{\bm{\theta}}) = \\
        &\sum_{seg,ifo} \Bigg[ \sum_{j=0}^{N-1} \log f_{seg,ifo} 
        \left(g_{seg,ifo}(\omega_j, \mu, \bm{\theta}), \tilde{\bm{\theta}}\right) \\
        & + \log\mathcal{N}_{seg,ifo}\left(\eta_\mathcal{R}(\omega_j)\right)\Bigg],
    \end{aligned}
\end{equation}
where the subscript $seg$ denotes the data segment, $ifo$ the interferometer, $\mu = \Lambda_i^{-2}$ is proportional to the amplitude of the DM peak, $\omega_j$ is the frequency in the $j^{th}$ bin, $\eta_\mathcal{R}$ is the ratio between measured and true strain data as determined by LIGO calibration~\cite{sun2020} with a corresponding Gaussian pull term $\mathcal{N}$, and finally $f_{seg,ifo}$ is a skew normal distribution with parameters denoted by $\tilde{\bm{\theta}}$, with $\bm{\theta}$ left to describe the spline and calibration parameters. $g_{seg,ifo}$ describes the residuals between the data and the expected background, with terms to model DM and calibration effects:
\begin{equation}
    \begin{aligned}
        g_{seg,ifo} =\ &2\log \eta_\mathcal{R} \\
        & + \log PSD_{seg} \\
        & - \log\left(y_{spline} + \mu\beta_{ifo}\right),
    \end{aligned}
\end{equation}
where $y_{spline}$ refers to the fitted background shape, and $\beta_{ifo}$ is an interferometer-specific calibration term based on~\cite{goettel2024}. Frequency-dependence throughout the equation is left implicit for simplicity. One should note that this likelihood function assumes that the stochasticity of the DM signal~\cite{Nakatsuka:2022gaf,Centers2021} is negligible because, by design, our integration time always matches an entire coherence time.

% Introduce q_0
\noindent Within this framework, searching for DM is equivalent to rejecting the hypothesis that $\mu = 0$. We use the profile-likelihood-ratio based test statistic $q_0$ (as described in~\cite{cowan2011}) which assumes $\mu \geq 0$:
\begin{equation} \label{eq:q0}
    q_0 = \begin{cases}    
    -2 \ln\frac{\mathcal{L}(0, \hat{\hat{\bm{\theta}}})}{\mathcal{L}(\hat{\mu}, \hat{\bm{\theta}})} & \text{if } \hat{\mu} \geq 0 \\
    0 & \text{if } \hat{\mu} < 0
    \end{cases},
\end{equation}
where $\hat{\hat{\bm{\theta}}}$ are the values of $\bm{\theta}$ that maximise the likelihood for a given $\mu$ and ($\hat{\mu}$, $\hat{\theta}$) are the values that globally maximise the likelihood. $q_0$ is zero when $\hat{\mu} < 0$ because since DM cannot induce negative peaks in the PSD, one must not regard under-fluctuations in the data as being incompatible with the null hypothesis.
%Further, since the $\Tilde{\bm{\theta}}$ values are determined in advance and kept constant in the likelihood maximisation \textit{etc..} they are not shown in the equation.
%The validity of these approximations was tested against a full Monte-Carlo calculation for a few randomly selected frequencies.

% Introduce q_mu
\noindent Conversely, in order to determine upper limits, we adopted the test statistic ${q}_{\mu}$:%\Tilde
\begin{equation} \label{eq:qmu}
    q_{\mu} = \begin{cases}
    -2 \ln\frac{\mathcal{L}(\mu, \hat{\hat{\bm{\theta}}})}{\mathcal{L}(\hat{\mu}, \hat{\bm{\theta}})} & \text{if } \hat{\mu} \leq \mu \\
    0 & \text{if } \hat{\mu} > \mu
    \end{cases},
\end{equation}
with the same nomenclature as in \Cref{eq:q0}.
%Larger values of $q_{\mu}$ represent greater incompatibility between the data and a hypothesised value of $\mu$.
One should note the sign reversal: $q_{\mu}$ is set to zero when $\hat{\mu} > \mu$ because in the case of setting an upper limit, higher values of $\mu$ do not represent incompatibility with the hypothesis. %After testing the validity of the asymptotic formul\ae\ described in~\cite{cowan_asymptotic_2011} with a full Monte-Carlo treatment, we adopt them to calculate upper limits over the full frequency range. The results are described in \Cref{sec:results}
To convert computed values of $q$ to significance levels for rejecting the null hypothesis, we first selected a thousand logarithmically spaced frequency bins within our analysis range. We then calculated the significance using both Monte-Carlo sampling of our likelihood and asymptotic formul\ae\ as described in \cite{cowan2011}. Finding no statistically significant differences between the two methods, we opted for the latter (faster) one in all subsequent calculations.

\section{Results} \label{sec:results}
\begin{figure}[h]
    \centering
    \includegraphics[width=\linewidth]{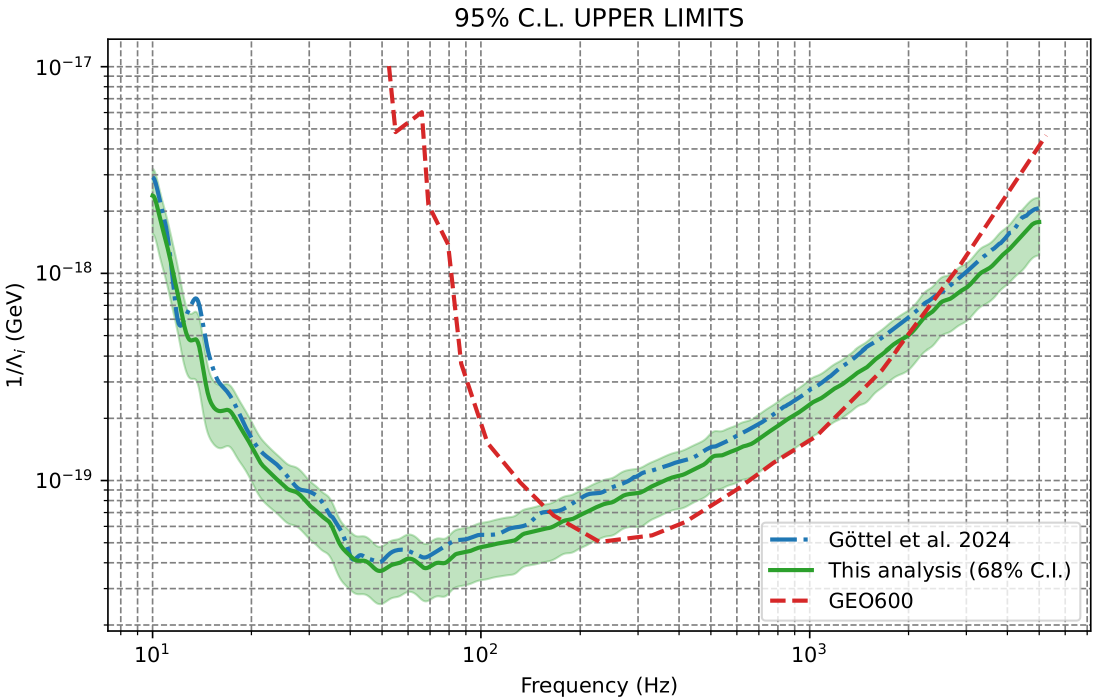}
    \caption{Upper limit on $\Lambda^{-1}_i$ (95\% C.L.) as a function of
frequency. The solid curve with a 68\% C.I. band represent the results obtained using spectral zero-suppression, the dot-dashed curve show the results from the band approximation (we note that the uncertainties are near-identical for both curves and were only shown once to help visualise the result), and the dashed line show the results from the GEO600 experiment~\cite{vermeulen2021a}. All curves are heavily smoothed for visual purposes.}
    \label{fig:results}
\end{figure}

The data used in this study is identical to that used in~\cite{goettel2024}, meaning 40 data segments from the third science run of the Hanford and Livingston detectors~\cite{OpenData}, selected as those in LIGO's third science run with a continuous measurement length of at least $\tau_c(\SI{10}{\hertz}) = \SI{27.78}{\hour}$. This data is calibrated from \SIrange{10}{5000}{\hertz}, which is why our results are evaluated in that range. We assume a local DM density of $\rho_{\rm local} = \SI{0.4}{\giga\electronvolt\per\cubic\centi\meter}$~\cite{read2014}.
%Because a search using the \textit{band approximation} was already performed on the same data, it was not repeated here.
We used the strategy described in~\cite{goettel2024} to find DM candidates on data obtained using the \textit{zero suppression} method, resulting in zero candidates. The resulting upper limits, from both methods, can be seen in \Cref{fig:results}. As expected, the spectral zero-suppression method shows a consistent improvement ($15\%$ when averaged over all frequencies), as it avoids the approximations required by the band approximation method. It also avoids the need for additional post-processing which would introduce nuisance degrees of freedom to the analysis.
In terms of computational cost, both methods are similar. Specifically, the zero-suppression method was an average of $46\%$ faster on segments shorter than \SI{36.4}{\hour}, but $34\%$ slower on longer segments. This is because at LIGO's sampling rate of \SI{16384}{\hertz}, the number of data points in these latter segments exceeds $2^{31}$. As a result, the length of $\Tilde{X}$ (see \Cref{sec:constQ}) suddenly doubles, leading to a corresponding increase in the number of operations required in the Fourier domain. Overall, the comparable cost of both methods highlights the spectral zero-suppression method's ability to recover FFT properties.

\section{Discussion}
% Optimal and fast
% Full potential not fully achieved, but already enough to beat my previous method
% Can be used not only for scalar, but for all DM searches
% If used, will maximise the scientific output of future detectors
% 
This work unveils an innovative method that, for the first time, achieves optimal scalar dark matter search results while keeping computational costs in check. By using insights from computer music analysis~\cite{brown1992a}, this approach does not only surpass a previous method, that relied on approximations~\cite{goettel2024}, in terms of results and efficiency, but it also demonstrates versatility. Indeed, the zero-suppression threshold applied here (see \Cref{sec:constQ}) was stringent, and a more relaxed approach could re-introduce some losses while reducing computational costs by at least another order of magnitude. Future work could fine-tune this balance to meet specific targets. Such high-speed capabilities would be invaluable for testing and diagnostics during development.

\noindent Moreover, while our study focused on a particular application, our method's potential extends beyond scalar dark matter searches. It can in principle be applied to any search involving signals with widths proportional to their frequency, encompassing most gravitational-wave detector-based dark matter searches~\cite{morisaki2021,abbott2022,LIGOVectorDM,Miller:2020vsl,Miller:2023kkd,Miller:2022wxu}. Ultimately, this approach has the potential of maximizing the scientific yield of future gravitational-wave detectors, both on Earth and in space~\cite{ET,CE,LISA,Taiji,TianQin}.

\begin{acknowledgments}
This research has made use of data or software obtained from the Gravitational Wave Open Science Center (gwosc.org), a service of the LIGO Scientific Collaboration, the Virgo Collaboration, and KAGRA. This material is based upon work supported by NSF's LIGO Laboratory which is a major facility fully funded by the National Science Foundation, as well as the Science and Technology Facilities Council (STFC) of the United Kingdom, the Max-Planck-Society (MPS), and the State of Niedersachsen/Germany for support of the construction of Advanced LIGO and construction and operation of the GEO600 detector. Additional support for Advanced LIGO was provided by the Australian Research Council. Virgo is funded, through the European Gravitational Observatory (EGO), by the French Centre National de Recherche Scientifique (CNRS), the Italian Istituto Nazionale di Fisica Nucleare (INFN) and the Dutch Nikhef, with contributions by institutions from Belgium, Germany, Greece, Hungary, Ireland, Japan, Monaco, Poland, Portugal, Spain. KAGRA is supported by Ministry of Education, Culture, Sports, Science and Technology (MEXT), Japan Society for the Promotion of Science (JSPS) in Japan; National Research Foundation (NRF) and Ministry of Science and ICT (MSIT) in Korea; Academia Sinica (AS) and National Science and Technology Council (NSTC) in Taiwan.
The authors are grateful for computational resources provided by the LIGO Laboratory and Cardiff University and supported by National Science Foundation Grants PHY-0757058 and PHY-0823459, and STFC grants ST/I006285/1 and ST/V005618/1. This material is based upon work supported by NSF's LIGO Laboratory which is a major facility fully funded by the National Science Foundation.
\end{acknowledgments}

\appendix
\section{} \label{app:analyticalKernel}
The time kernel we wish to analytically Fourier transform is the product of our window function and an exponential term, as defined in \Cref{eq:defineKernel}. In this study, we use the Kaiser-Bessel window~\cite{nuttall1981}, defined as:
\begin{equation}
    w(x) = \frac{1}{L I_0(\beta)}I_0\left(\beta\sqrt{1 - (2x/L)^2}\right) \text{for}\ |x| < L/2,
\end{equation}
where the window is of length $L$ and $\beta$ is a scaling parameter that allows some level of control over main and side lobe levels. This window was designed to be an analytical close-to-optimum solution to maximise total power in the main lobe. We use $\beta=238.13$ (based on~\cite{vermeulen2021a}). The discrete Fourier transform of the Kaiser-Bessel window is a complex operation, but can be solved for the real part:
\begin{equation} \label{eq:analyticalKernel}
    \mathcal{F}(\Re(w))(f) = 
    \begin{cases} 
    \frac{\sin{\sqrt{(\pi L f)^2 - \beta^2}}}{I_0(\beta)\sqrt{(\pi L f)^2 - \beta^2}}, & (\pi L f)^2 - \beta^2 \geq 0 \vspace{.2cm} \\
    \frac{\sinh{\sqrt{\beta^2 - (\pi L f)^2}}}{I_0(\beta)\sqrt{\beta^2 - (\pi L f)^2}}, & \text{otherwise},
    \end{cases}
\end{equation}
where $f$ is the frequency. The two cases above are mathematically equivalent but shown separately to reflect software implementation. This solution lets us easily compute the (real part of the) spectral kernels, because the aforementioned exponential term can simply be interpreted as a discrete frequency-shift in Fourier space. In order to use these results however, two conditions must be fulfilled. Firstly, for \Cref{eq:analyticalKernel} to be valid, the time kernel must be real. This means:
\begin{align}
    \Im\left(e^{-2\pi i n \mathcal{Q} / N(j)}\right) &= 0 \nonumber \\
    \leftrightarrow -2\pi n \mathcal{Q} / N(j) &= 0\mod\pi,
\end{align}
and since both $n$ and $N(j)$ are integers, the relation holds if $\mathcal{Q}$ is an integer too. Given the high value of $\mathcal{Q}$ ($\approx 10^6$), this can be set as a negligible approximation. Secondly, since \Cref{eq:analyticalKernel} only contains real terms, we must ensure that the spectral kernel is real, \textit{i.e.} that the time kernel is conjugate symmetric. We achieve this by following the approach in~\cite{brown1992a}, implementing a discrete version of the window that is symmetric around the centre of the interval. Assuming an odd window length $N$ this can be written as:
\begin{align} \label{eq:symmetric_window}
    w^\prime(n) &= w(n - (N-1)/2), \nonumber \\
    &= \begin{cases}
        \frac{I_0\left(\beta\sqrt{1 - \left(\frac{2n}{N - 1} - 1\right)^2}\right)}{I_0(\beta)} & \text{if } 0 \leq n \leq N \\
        0 & \text{otherwise}.
    \end{cases}
\end{align}
Applying the same transformation $n\rightarrow n - (N-1)/2$ to the exponential term of the time kernel results in a conjugate symmetric time kernel. This, combined with the integer requirement on $\mathcal{Q}$, allows us to directly evaluate our spectral kernel.
%In the analysis, this transformation is applied on the spectral side, which is achieved trivially by inserting the terms in \Cref{eq:analyticalKernel}.
However, it also induces the need for an \textit{a posteriori} correction to align the time domain data as required by our PSD calculation in \Cref{eq:DFT}. This can be understood with the sketch below:
\begin{tikzpicture}
    \def\lineEND{\linewidth*0.9}
    % Main horizontal line
    \draw[thick] (0,0) -- (\lineEND*3/4,0);
    \draw[dashed] (\lineEND*3/4,0) -- (\lineEND,0);
    \draw[thick] (0, -0.1) -- (0, 0.1);
    \draw[thick] (\lineEND, -0.1) -- (\lineEND, 0.1);
    \node[above] at (\lineEND/2, 0.1) {$N_\text{FFT}$/2};
    \node[above] at (\lineEND, 0.1) {$N_\text{FFT}$};

    % Inset horizontal line
    % \draw[thick] (0,-0.5) -- (\linewidth*3/4, -0.5);
    \draw[thick] (\lineEND*3/4, -0.1) -- (\lineEND*3/4, 0.1);
    \node[above] at (\lineEND*3/4,0.1) {$N_\text{segment}$};

    % Smaller horizontal line
    \draw[thick] (\lineEND*0.3,-.3) -- (\lineEND*0.7,-.3);
    \draw[thick] (\lineEND*0.3, -.4) -- (\lineEND*0.3, -.2);
    \draw[thick] (\lineEND*0.7, -.4) -- (\lineEND*0.7, -.2);
    \node[below] at (\lineEND/2, -.3) {$N_\text{section}$};
    \draw[densely dotted] (\lineEND/2, 0.1) -- (\lineEND/2, -0.6);

    % Diff line
    \def\depth{-.6}
    \draw[thick] (0, \depth) -- (\lineEND*0.3, \depth);
    \draw[thick] (0, \depth-.1) -- (0, \depth+.1);
    \draw[thick] (\lineEND*0.3, \depth-.1) -- (\lineEND*0.3, \depth+.1);
    \node[below] at (\lineEND*0.15, \depth+.1+.4) {$\delta_t$};
    % \node[below] at (\lineEND*0.15, \depth-.1) {($N_{FFT}$ - $N_\text{section}$)/2};
\end{tikzpicture}

\noindent where (NOT to scale) $N_\text{segment}$ is the number of data points in a given strain segment, $N_\text{FFT}$ is the lowest power of two that is equal or higher than $N_\text{segment}$, and $N_\text{section}$ is the frequency-dependent length of the sections being summed over (see \ref{sec:constQ}). The kernel (see \Cref{eq:symmetric_window}) now defined as symmetric around the centre of the interval, which is in practice defined by $N_{\text{FFT}}$, would describe the data under the line labelled $N_\text{section}$. It is possible to align the window to the beginning of our data segment by shifting it by $\delta_t = (N_{FFT} - N_\text{section})/2$. We achieve this by multiplying the previously-obtained spectral kernel by a corresponding $e^{2 \pi i n \delta_t}$ term. By shifting the time kernel through this retroactive multiplication of the spectral kernel, we can keep its necessary complex-conjugate properties as derived above during the Fourier transform step, but are still able to implement the sum over sections required by \Cref{eq:PSD}.

\noindent Inserting the steps just described into the calculation of the power terms (right side of \Cref{eq:defineKernel}), for any segment $s \in [0, ..., N_\text{sections}]$, results in:
\begin{equation} \label{eq:Xjs}
    % X_{j,s} = \frac{1}{N}&\sum_{j=0}^{N(j)-1} \Tilde{X}_j \Tilde{K}_j(j/N_{FFT} - \mathcal{Q}/N(j)) \cdot\nonumber\\
    % &\exp\left(\frac{2\pi i j}{N_{FFT}} \left(\frac{1}{2}(N_{FFT} - N(j)) - s\cdot\delta_s(j)\right)\right),\\
    X_{j,s} = \frac{1}{N}\sum_{j=0}^{N(j)-1}\Tilde{X}_j\cdot\Tilde{K}_{j^\prime}\cdot e^{\frac{2\pi i j}{N_{FFT}} N^\prime(j, s)} \nonumber,
\end{equation}
where:
\begin{align}
    j^\prime &= j / N_\text{FFT} - \mathcal{Q} / N(j) \nonumber \\
    N^\prime(j, s) &= \frac{N_\text{FFT} - N(j)}{2} - s\delta_s(j),
\end{align}
with $\delta_s(j) = (1 - \xi)\cdot N(j)$ as the distance between data sections for a given overlap $\xi$. The $j/N_{FFT} - \mathcal{Q}/N(j)$ term stems from interpreting the exponential component in the time kernel as a discrete shift in frequency space, the $(N_{FFT} - N(j))/2$ term makes use of the same discrete shift, but in time space, to ensure that the summation over the windows correctly starts at index 0. Finally, the $s\delta_s(j)$ term is used to shift the data over the different sections to perform averaging as detailed in \Cref{eq:PSD}.

\section{} \label{app:normalisation}
The calculation of the total power in a given frequency bin, see \Cref{eq:PSD}, requires a normalisation term~\cite{troebs2006} in which a full window function must be evaluated in each frequency bin. Since we are using a Kaiser-Bessel window function, each of those terms requires a call to the modified Bessel function of zeroth order. This is a slow operation, and while its cost was bearable in past analyses, given the speed-up factors achieved in this study, it becomes a bottleneck. We thus strive to calculate this term analytically and begin by setting:
\begin{equation} \label{eq:norm}
    \sum_{n=0}^{N_j - 1} w_n^2 \approx \frac{1}{I_0^2(\beta)}\int_0^{N_j}dn\ I_0^2\left(\beta\sqrt{1 - \left(\frac{2n}{N_j-1} - 1\right)^2}\right).
\end{equation}
\noindent This is a very natural approximation given the integration lengths ($10^{5-9}$) in our analysis. We can then use the summation notation of the Bessel function to take the Cauchy product of $I_0^2$:

\begin{align}
    I_0^2(x) &= \sum_{k=0}^\infty\sum_{l=0}^k\frac{1}{((l!)(k - l)!)^2}\cdot\left(\frac{x}{2}\right)^{2k} \nonumber \\
    &= \frac{1}{\sqrt{\pi}}\sum_{k=0}^\infty\frac{x^{2k}}{(k!)^3}\Gamma\left(\frac{2k+1}{2}\right).
\end{align}
\vfill
\noindent Implementing this result into \Cref{eq:norm} and shifting the integral inside of the sum, we continue with:

\begin{align}
    &= \frac{1}{\sqrt{\pi}} \sum_{k=0}^\infty \frac{\Gamma\left(\frac{2k+1}{2}\right)}{(k!)^3}\int_0^{N_j} dn\ \beta^{2k}\left(1 - \left(\frac{2n}{N_j} - 1\right)^2\right)^k \nonumber \\
    &= N_j\frac{1}{\sqrt{\pi}}\sum_{k=0}^\infty  \frac{\Gamma\left(\frac{2k+1}{2}\right)}{(k!)^3} (2\beta)^{2k} \mathcal{B}(k+1, k+1),
\end{align}

\noindent where $\mathcal{B}$ is the beta function, and we have omitted the $I_0^2(\beta)$ term for simplicity. While this can be evaluated numerically, we find that the most important result here is that for a given $\beta$, this integral simplifies to be fully proportional to $N_j$. In this analysis, we can thus calculate the full normalisation for a single frequency bin, and then simply scale by multiplication with $N_j$.
\vfill

\vfill

% \bibliography{methods}% Produces the bibliography via BibTeX.

\end{document}